\documentclass[twocolumn,showpacs,pra,amsmath,amssymb]{revtex4}
\usepackage{epsfig}
\usepackage{epsfig,amsmath}
\usepackage{graphicx}
\usepackage{dcolumn}

\newcommand{\beq}{\begin{equation}}
\newcommand{\eeq}{\end{equation}}
\newcommand{\beqa}{\begin{eqnarray}}
\newcommand{\eeqa}{\end{eqnarray}} 
 \newcommand{\rh}{\rho}

\def\ejp#1{{\it  Eur.\ J.\ Phys.} {\bf #1}}

\def\oc#1{{\it  Opt.\ Commun.} {\bf#1}}
\def\ol#1{{\it  Opt.\ Lett.} {\bf#1}}

\def\jpb#1{{\it  J.\ Phys.\ B} {\bf#1}}
\def\jpa#1{{\it  J.\ Phys.\ A} {\bf#1}}
\def\nat#1{{\it  Nature} {\bf#1}}
\def\natp#1{{\it  Nature\ Phys.} {\bf#1}}

\def\pr#1{{ \it Phys.\ Rev. } {\bf#1}}
\def\pra#1{{\it  Phys.\ Rev. A\/} {\bf#1}}
\def\prb#1{{\it  Phys.\ Rev. B\/} {\bf#1}}

\def\prl#1{{\it  Phys.\ Rev.\ Lett.} {\bf#1}}
\def\qic#1{{\it  Quant.\ Inf.\ Comp.} {\bf#1}}
\def\sci#1{{\it  Science} {\bf#1}}
\def\rmp#1{{ \it Rev. \ Mod. \ Phys.} {\bf#1}}
\def\natp#1{{\it  Nature\ Phys.} {\bf#1}}

\begin{document}

\title{Sudden Death of Entanglement}

\author{Ting Yu$^1$\footnote{Email address:
ting.yu@stevens.edu} and J.~H. Eberly$^2$\footnote{Email address:
eberly@pas.rochester.edu}} \affiliation{$^1$Department of Physics and Engineering Physics,
Stevens Institute of Technology,
Hoboken, New Jersey 07030-5991, USA\\
$^2$Rochester Theory Center
and Department of Physics and Astronomy, University of Rochester,
Rochester, New York 14627-0171, USA}
\date{\today}

\begin{abstract}

 A new development in the dynamical behavior of elementary quantum systems is the surprising discovery that correlation between two  quantum units of information called qubits can be degraded by environmental noise in a way not seen previously in studies of dissipation. This new route for dissipation attacks quantum entanglement, the essential resource for quantum information as well as the central feature in the Einstein-Podolsky-Rosen so-called paradox, and in discussions of the fate of Schr\"odinger's Cat. The effect has been labelled ESD, standing for early stage disentanglement or, more frequently, entanglement sudden death. We review recent progress in studies focused on this phenomenon.
\end{abstract}

\maketitle

 A new development in the dynamical behavior of elementary quantum systems is the surprising discovery that correlation between two  quantum units of information called qubits can be degraded by environmental noise in a way not seen previously in studies of dissipation. This new route for dissipation attacks quantum entanglement, the essential resource for quantum information as well as the central feature in the Einstein-Podolsky-Rosen so-called paradox, and in discussions of the fate of Schr\"odinger's Cat. The effect has been labelled ESD, standing for early stage disentanglement or, more frequently, entanglement sudden death. We review recent progress in studies focused on this phenomenon.
 
 \begin{figure}[!b]
\includegraphics[width = 6cm]{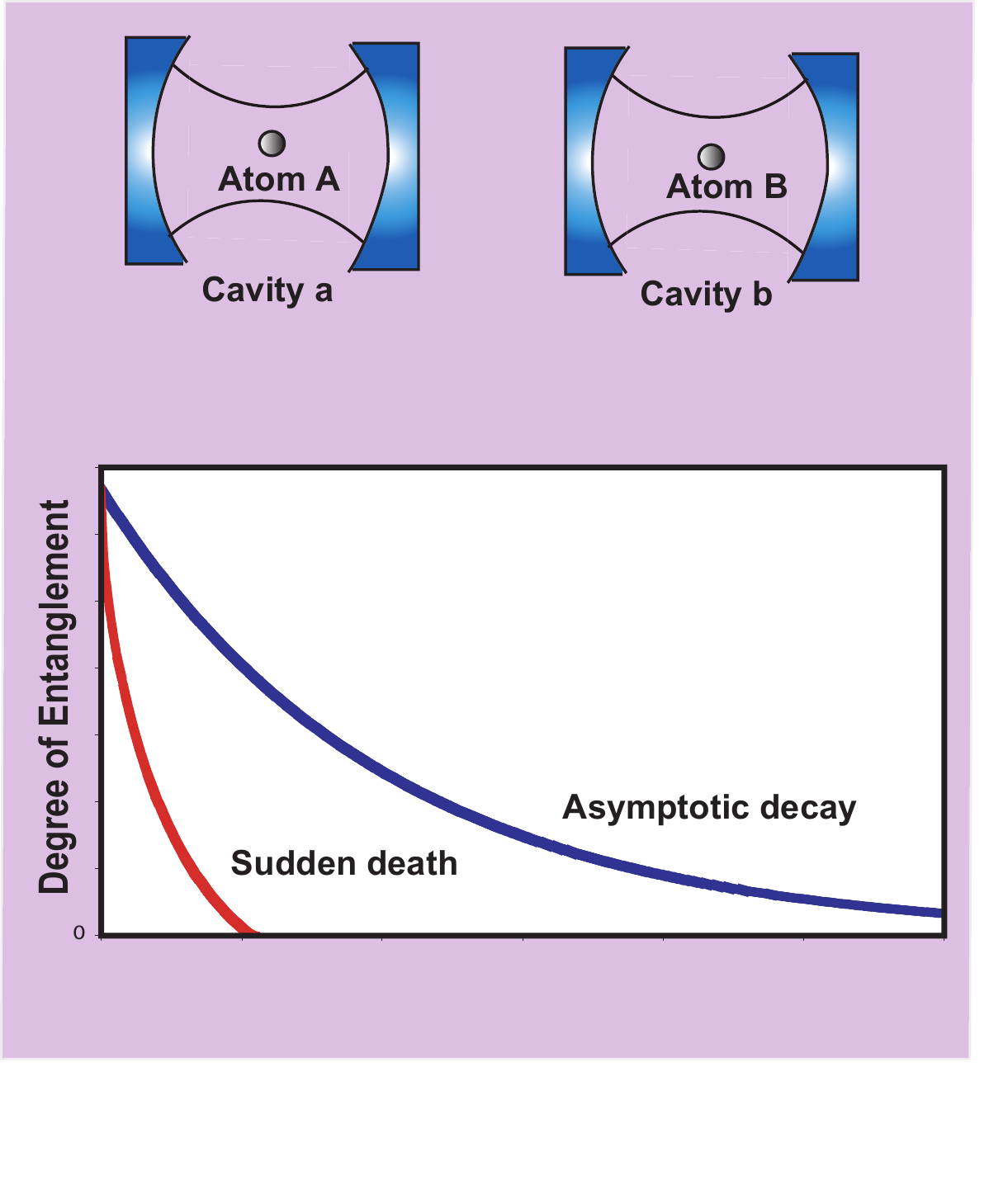}
\caption{ \label{scifig1} Curves clearly show ESD as one of two routes for relaxation of the entanglement, via concurrence $C(\rho)$, of qubits $A$ and $B$ that are located in separate overdamped cavities (adapted from ({\it 10})) }
\end{figure}

Quantum entanglement is a type of correlation, but special because it can be shared only among quantum systems. It has been the focus of foundational discussions of quantum mechanics since the time of Schr\"odinger (who gave it its name) and the famous EPR paper of Einstein, Podolsky and Rosen \cite{Schrodinger, EPR}. The degree of correlation available with entanglement is predicted to be stronger as well as qualitatively different compared to any other known type of correlation. Entanglement may be highly non-local, e.g., shared among pairs of atoms, photons, electrons, etc., even though remotely located and not interacting with each other, and these features have recently promoted the study of entanglement as a resource that we believe will eventually be put to use in new approaches to both computation and communication, for example by improving previous limits on speed and security, in some cases dramatically \cite{Nielsen-Chuang, Bennett-DiVincenzo}.

Quantum and classical correlations alike always decay as a result of  noisy backgrounds and decorrelating agents that reside in ambient environments \cite{Zurek03}, so the degradation of entanglement shared by two or more parties is unavoidable \cite{Yu-Eberly02,Simon-Kempe02, Dur-Briegel04, report}. The background agents with which we are concerned  have extremely short (effectively zero) internal correlation times themselves and their action leads to the familiar law mandating that after each successive half-life of decay there is still half of the prior quantity remaining, so that a diminishing fraction always remains. 

However, a theoretical treatment of two-atom spontaneous emission \cite{Yu-Eberly04PRL} shows that quantum entanglement doesn't always play the game according to the half-life law. Earlier studies of two-party entanglement in different model forms also pointed to this fact \cite{Zyczkowski-etal, Diosi, Rajagopal-Rendell, Daffer-etal, Dodd-Halliwell}. The term now used, entanglement sudden death (ESD, equally reasonably called early-stage disentanglement), refers to the fact that even a very weakly dissipative environment can degrade the specifically quantum portion of the correlation to zero in just a finite time (Fig. 1) rather than by successive halves. We will use the term decoherence to refer to the loss of quantum correlation, loss of entanglement.

This finite-time dissipation is a new form of decay  \cite{Yu-Eberly06PRL}, predicted to attack only quantum entanglement, and not previously encountered in the dissipation of other physical correlations. It has been found in numerous theoretical examinations to occur in a wide variety of entanglements involving pairs of atomic, photonic and spin qubits, continuous Gaussian states, and subsets of  multiple qubits and spin chains \cite{SOM1}. ESD has already been detected in the laboratory in two different contexts \cite{Almeida-etal07, Laurat-etal}, confirming its experimental reality and supporting its universal relevance \cite{Eberly-Yu07Sci}. Despite this, there is still no deep understanding of sudden death dynamics, and so far there is no generic   preventive measure.

\section*{How does entanglement decay?}
An example of an ESD event is provided by the weakly dissipative
process of spontaneous emission, if the dissipation is ``shared" by
two atoms (Fig. 1). To describe this we need a suitable notation.

The pair of states for each
atom, sometimes labelled $(+)$ and $(-)$ or sometimes (1) and (0),
are quantum analogs of ``bits" of classical information,
and for this reason such atoms (or any quantum systems with just two states) are called quantum bits or ``qubits". Unlike classical bits, the states of the atoms have the quantum ability to exist in both states at the same time. This is the kind of superposition used by
Schr\"odinger when he introduced his famous Cat, neither dead nor
alive but both, in which case the state of his Cat is conveniently coded by the bracket $[+ \Leftrightarrow -]$, in order to indicate equal simultaneous presence of the opposite $+$ and $-$ conditions.

This bracket notation can be extended to show entanglement. Suppose we have two opposing conditions for two cats, one large, one small and either waking (W) or sleeping (S). Entanglement of idealized cats could be denoted with a bracket such as $[(Ws) \Leftrightarrow (Sw)]$, where we've chosen large and small letters to distinguish a big cat from a little cat. The bracket would signal via the term $(Ws)$ that the big cat is awake and the little cat is sleeping, but the other term $(Sw)$ signals that the opposite is also true, big cat sleeping and little cat awake. 

One can see the essence of entanglement here: If we learn that the big cat is awake, the $(Sw)$ term must be discarded as incompatible with what we learned and so the two-cat state reduces to $(Ws)$. We immediately conclude that the little cat is sleeping. Thus knowledge of the state of one of the cats conveys information about the other \cite{EntOpposites}. The brackets are symbols of information about the cats' states, and don't belong to one cat or the other. The brackets belong to the reader, who can make predictions based on the information the brackets convey. The same is true of all quantum mechanical wave functions.

Entanglement can be more complicated, even for idealized cats. Then a two-party joint state must be represented not by a bracket as above, but by a matrix, called a density matrix and denoted $\rho$ in quantum mechanics (see \cite{SOM2} and Eqn. (S3)). When exposed to environmental noise, the density matrix $\rho$ will change in time, becoming degraded, and the accompanying change in entanglement can be tracked with a quantum mechanical variable called concurrence \cite{Wootters}, which is written for qubits such as the atoms $A$ and $B$ in Fig. 1 as:
\beq \label{concurrence}
C(\rh) = \max[\ 0,\ Q(t)\ ],
\eeq
where $Q(t)$ is an auxilliary variable defined in terms of entanglement of formation, as given explicitly in Eqn. (S4).  $C=0$ means no entanglement and is achieved whenever $Q(t)$ is negative,  while for maximally entangled states one has $C=1$, and $C$ is limited to that range: $1 \ge C \ge 0$.

In the case of spontaneous emission there is no environment at all, except for the vacuum. The vacuum can still have a noisy degrading effect through its quantum fluctuations, which can't be avoided, so both atoms in Fig. 1 must eventually lose their excitation and come to their ground states. Then their state is simply $(--)$, a completely disentangled situation because learning that the state of one is $(-)$ doesn't change our information about the other, also $(-)$. Thus disentanglement is the eventual fate of the pair. 

The question is, how quickly do they meet their fate? For the initial density matrix shown in Eqn. (S5),  the answer is supplied by the surface graphed in Fig. 2, which shows  possible pathways for entanglement dissipation as a function of time. The $\kappa$  axis shows that the time evolution of entanglement depends on the value of the parameter  that encodes the initial probability for the two atoms to be in the doubly excited state $(++)$. The two extreme concurrence curves for $\kappa=1$ and $\kappa=0$ are the ones already shown in Fig. 1. 

\begin{figure}[!b]
\includegraphics[width = 6cm]{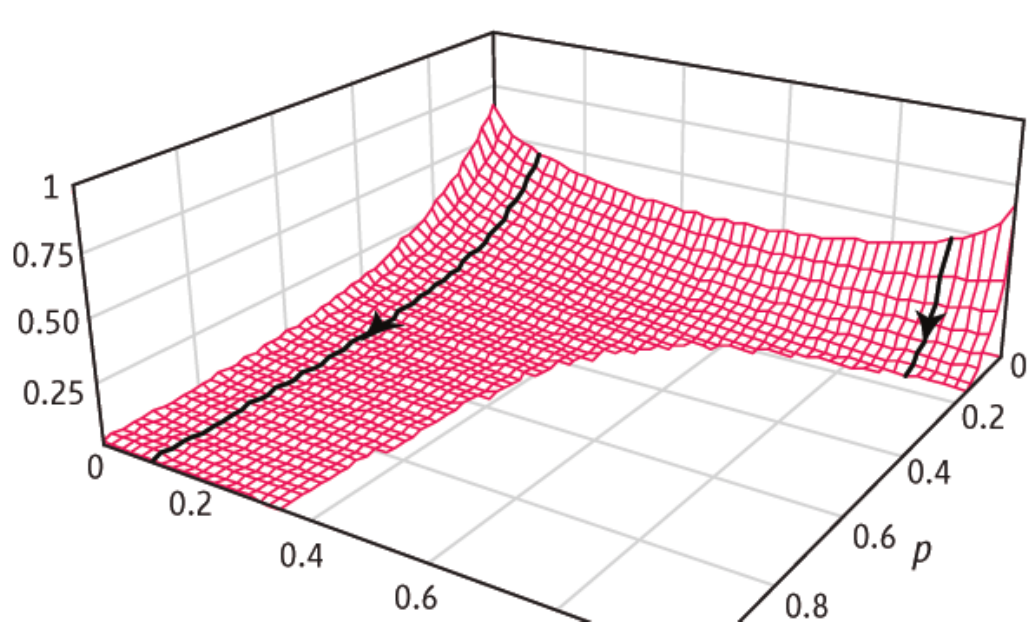}
\caption{ \label{scifig2}  Atom-atom entanglement is plotted as a function of time for $\kappa$ values in the range 0-1 (adapted from ({\it 20})). For all values of $\kappa$ less than 1/3, the half-life rule is obeyed, but for $\kappa$ between 1/3 and 1 it is not. For those values the curves show ESD, i.e., becoming zero in a finite time and remaining zero thereafter. The two curves marked with arrows are similar to the curves in Fig. 1.  The time is represented by $p=1-\exp(-\Gamma t)$.}
\end{figure}

\begin{figure}[!t]
\includegraphics[width=6cm]{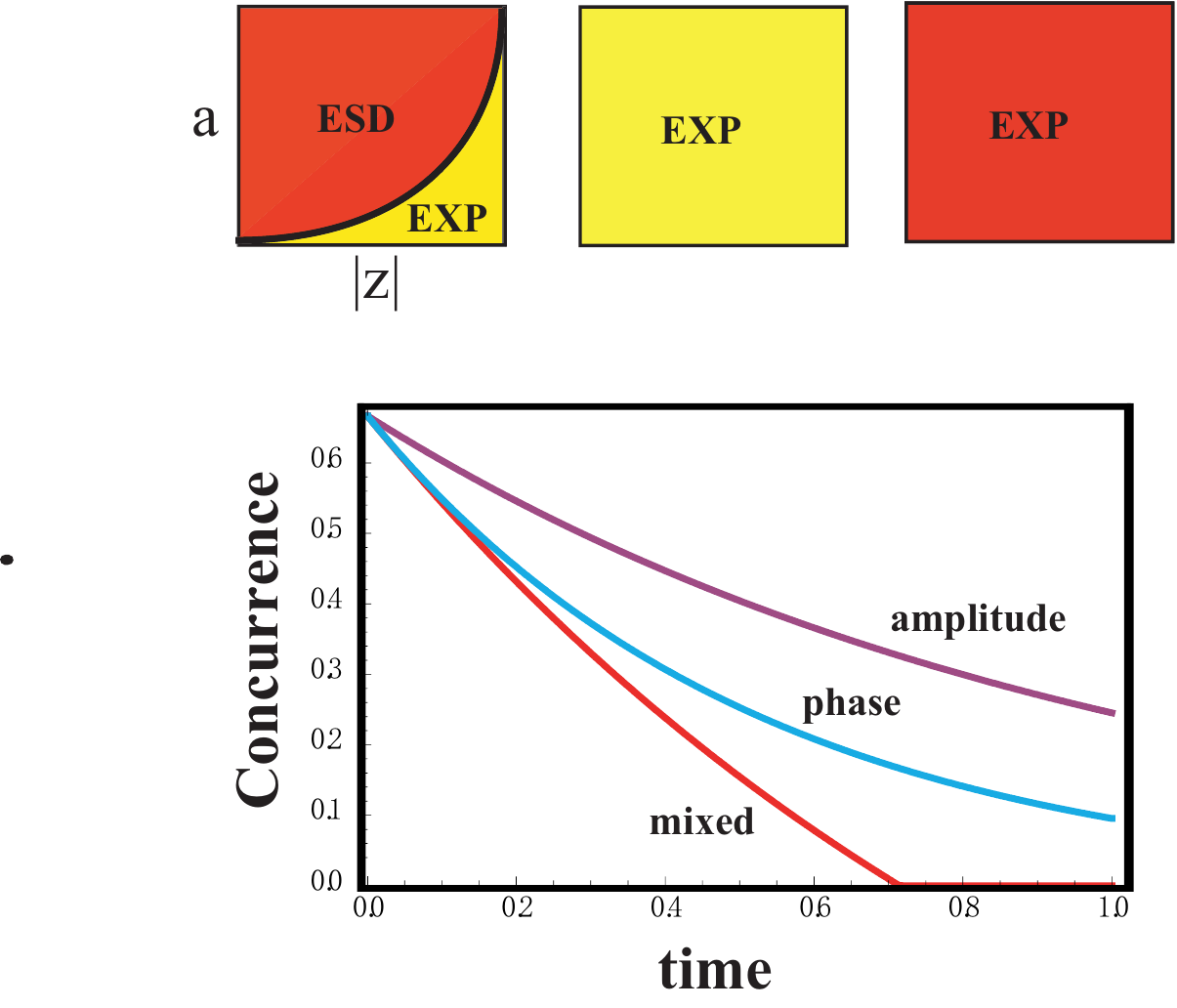}
\caption{ \label{scifig3}  The first two time-dependent curves show exponential (smooth half-life-type) decay of concurrence for a qubit pair exposed to phase noise and to amplitude noise, respectively. The bottom curve shows non-smooth decay, i.e., ESD occurs for the qubit pair when both noises are acting together. The color-coded squares apply to any two-party $X$ matrix (S3) having $d=0$. They present the predicted results for the entire accessible physical domain, which means throughout $1 \ge a \ge 0$ and $1 \ge |z| \ge 0$. The yellow zones labelled EXP designate domains where smooth exponential (half-life type) evolution occurs and the red zones show where ESD occurs. The first two squares apply when amplitude and phase noises are applied separately, and have smooth decay regions, while the final entirely red square shows that sudden death is universal in the entire region for any $X$ matrix (S3). }
\end{figure}

The sudden death behavior shown in the right highlighted curve of Fig. 2 is a new feature for physical dissipation \cite{Yu-Eberly06PRL, Eberly-Yu07Sci}, and is induced by classical as well as quantum noises \cite{Yu-Eberly06OC}. It is counter-intuitive, based on all previous single-atom experience, because spontaneous emission is a process that obeys the half-life rule rigorously for individual atoms. But it turns out that the two-qubit correlation doesn't follow the one-qubit pattern. Said another way, the sudden death doesn't come from a shorter half-life; the entangled joint correlation doesn't even have the half-life property.

As reported in \cite{Almeida-etal07}, the first experimental confirmation of ESD was made with an all-optical approach focusing on photonic polarization. It was achieved by the tomographic reconstruction of $\rho(t)$, and from it the $Q(t)$ variable, and thus the concurrence $C(\rho)$. In the experiment both amplitude and phase noises that can degrade entanglement were realized by combining beam splitters and mirrors.

\section*{Can sudden death be avoided or delayed?}

The issue how to avoid ESD-type decorrelation in a realistic physical
system is incompletely resolved at this time. A number of methods are
known to provide protection against previously known types of
decorrelation \cite{Nielsen-Chuang}. Some methods have classical analogs in information theory. One engages appropriately designed redundancy, and is known as quantum error correction. Another relies on using a symmetry that can isolate entanglement from noise, effectively providing a decoherence-free subspace to manage qubit evolution.

It is known that error correction is most useful when the disturbing noise is below some threshold \cite{ErrorCorrP}. In practice error correction can be complicated, because a noisy channel is a dynamical process and its physical features are often not fully understood or predictable. An example is atmospheric turbulence during open-air communication. Another issue is the cost associated with providing redundancy. Additionally, it has been reported that some quantum error correction algorithms could actually promote rather than mitigate ESD \cite{Sainz-Bjork08}. Symmetries that avoid decoherence by providing isolation from noise during evolution \cite{Lidar-Whaley} have also been examined as a way to postpone or avoid ESD, but knowledge of the noise to be combatted appears unlikely to be available because qubits remote from each other would rarely share either symmetry properties or noise descriptions.  

Other methods considered use dynamic manipulation such as mode modulation \cite{Gordon-Kurizki} or the quantum Zeno effect \cite{Zeno08}, which can be regarded as extensions of the so-called bang-bang method \cite{Bang-Bang}. Another proposal is to use feedback control \cite{Carvalho-etal08} to prevent ESD and other decoherence effects in the presence of hostile noise. None of these methods is perfect, but more effective if designed for specific noise avoidance.

\section*{Does the number of noises matter?}

Nonlocal entanglement raises the issue of ESD triggered by different noise processes and can refer to more than one noise source acting together on co-located entangled qubits, or to independent noise sources acting separately on remotely located members of a qubit pair. The rate of dissipation in the presence of several noise sources is normally the sum of the individual dissipation rates. More explicitly, if decay rates $\Gamma_1$ and $\Gamma_2$ come from the action of two distinct weak noises, then when the two noises are applied together to a physical system, the resulting  relaxation rate is simply given by the sum of the separate rates: $\Gamma_1 + \Gamma_2$.

However, such a long-standing result does not hold for entanglement decay. This was discovered \cite{Yu-Eberly06PRL} by examining entanglement evolution of a set of $X$-form mixed state matrices (see Eqn. (S3)) with $d=0$. Straightforward calculations for the entire class are illustrated by the diagrams in Fig. 3, devoted to the specific example of two qubits exposed together to amplitude noise as well as phase noise. The top two time-dependent curves show that the application of either noise separately allows long-running entanglement decay of the half-life type (no ESD). The bottom curve is different, as it hits zero in a finite time. That is, the combined effect of the two noises surely causes ESD. The caption explains the colored squares. This illustrates the ``super-vulnerability" of paired-qubit entanglement when attacked by  different independent noises. This result is universal in the sense that it continues to hold (see \cite{Yu-Eberly06PRL}) even if the two noises attack one of the qubits but not the other, and also if the two qubits are remotely attacked each by just one of the noises.

\section*{Is there an ``anti-ESD" or rebirth effect?}

Special circumstances are needed to see ``anti-ESD", the creation or rebirth of entanglement from disentangled states. Of course, by imposing the right interactions almost anything can be made to happen, but we are  concerned with evolution of joint information in a pure sense, and focus on two-party entanglement that evolves  without mutual interaction or communication. 

The same two-atom situation shown in Fig. 1 can be made relevant to anti-ESD. In solving for the surface of solutions plotted in Fig. 2 the cavities were taken as fully overdamped, so that any photon emitted by either atom was irreversibly absorbed by the walls, but they could also be treated as undamped mirror-like cavities, such as used in the Jaynes-Cummings (JC) model for light-atom interactions \cite{Jaynes-Cummings}. This situation produces a periodic sequence of perfect rebirths of atom $AB$ entanglement \cite{Yonac-etal06, Ficek-Tanas}. An early mathematical model of two-qubit evolution \cite{Zyczkowski-etal} can be interpreted as treating an underdamped cavity and also shows rebirths.

The panels in the top row of Fig. 4 show rebirth scenarios. They occur for states that are initially of the cat type, such that both atoms are excited and both are un-excited at the same time. The cat-type bracket  for them is:
\beq \label{e.PhiAtoms}
\Phi_\alpha = [(++)\cos\alpha  \Leftrightarrow (--)\sin\alpha ], \\
\eeq
where different values of sine and cosine produce the different curves in the figure (see \cite{Yonac-etal07}).

\begin{figure}[!b]
\includegraphics[width = 7cm]{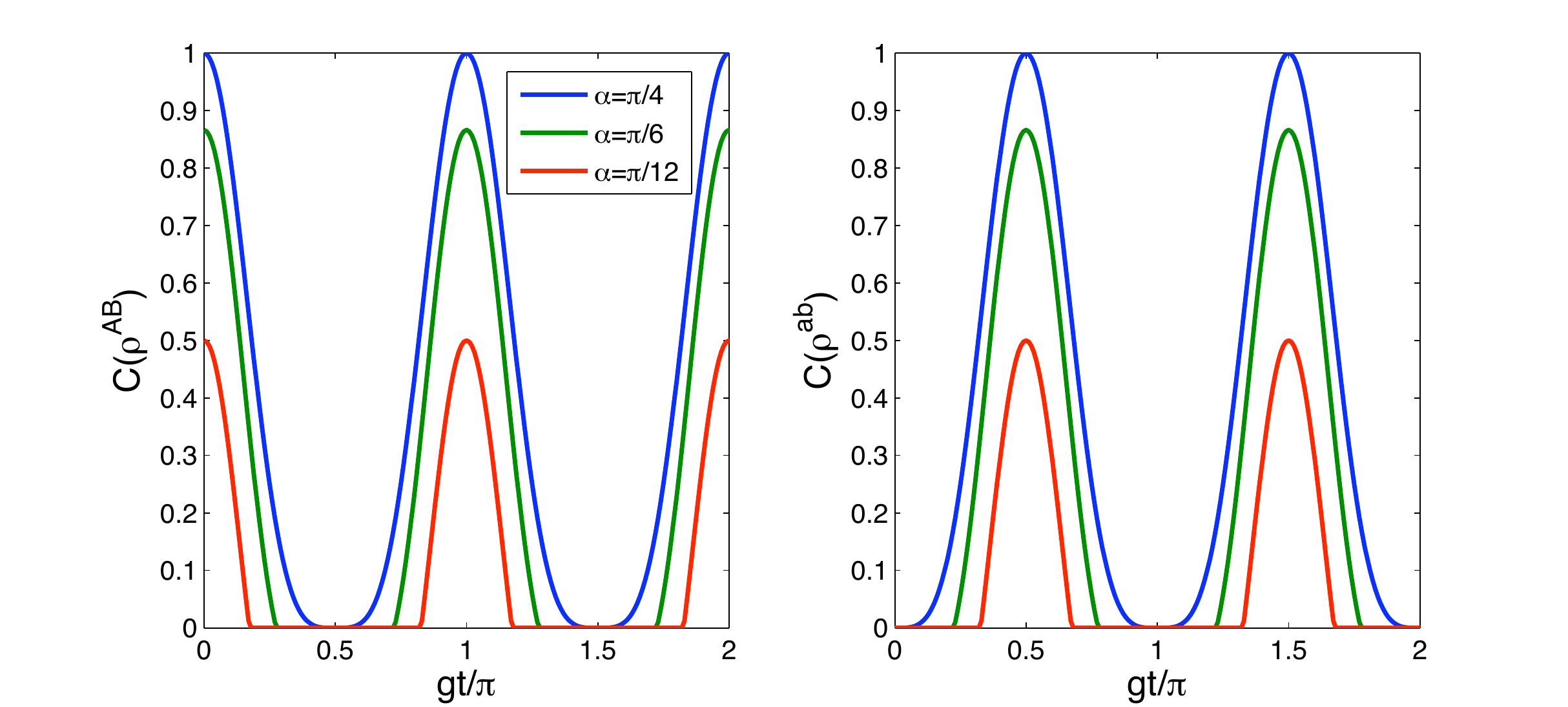}
\includegraphics[width =7cm]{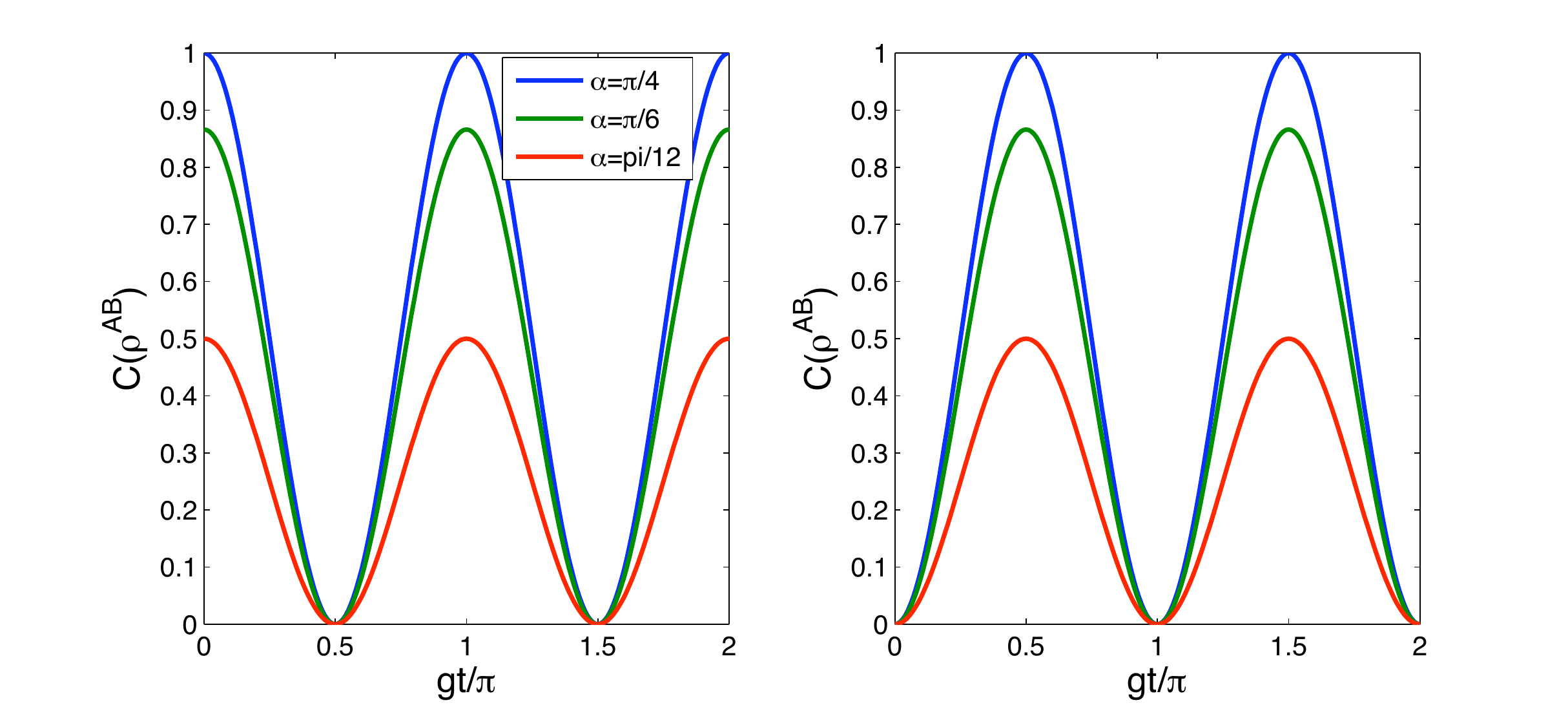}
\caption{ \label{scifig4}   Entanglement birth, death, and rebirth. [Adapted from (35)] In the bottom pair of panels the rise
and fall of AB entanglement is exactly compensated by the fall and rise of ab entanglement. This is not the
case in the top pair of panels, but a more subtle form of compensation still occurs, as reported in (35). It
involves not concurrence but the auxiliary variable $Q(t)$  defined in Eqs. 1 and S4. }
\end{figure}

Starting from the photon vacuum state in each cavity, the JC-type evolution will permit only zero or one photon to reside in each cavity at any later time, so each of the two modes is a two-state system (a qubit), and counting the two atoms there are now four qubits on hand. This provides six concurrences that can be computed: $C^{AB},\ C^{ab},\ C^{Ab},\ C^{aB},\ C^{Aa}$ and $C^{Bb}$, where the capital letters identify atoms and the small letters identify the photons in cavity modes $a$ or $b$. Concurrence is defined only for pairs of qubits, not quartets of them, so the label $C^{AB}$ implies that the $a$ and $b$ degrees of freedom are not available to observation and have been ignored (technically, have been traced out), and for photon concurrence $C^{ab}$ the atomic $A$ and $B$ properties are traced out, and so on.

This idealized model provides a convenient framework to analyze entanglement in a simplified but still multi-qubit framework. As shown in  the top half of Fig. 4, ESD takes place for atom concurrence (in the left panel almost all $C^{AB}$ curves hit zero and remain zero for finite times). However, in the right panel the photon concurrence $C^{ab}$ behaves oppositely to $C^{AB}$, showing anti-ESD. That is, initially $C^{ab}$ is zero, but it immediately begins to grow. The photons jointly experience entanglement ``sudden birth", but this is followed by ESD a half cycle later. All of this is via pure ``informatics", i.e., without energy exchange or other interaction between the sites.

The reason for the rebirths is obvious -- the photons emitted cannot get really lost among the few joint states available.  If a larger number of cavity modes would be provided, a longer time would be needed for a rebirth to be complete, and as a limiting case, the cavities producing the curves in Fig. 2 have an infinite number of modes, so the lost quantum correlation cannot be reborn in any finite time. If there are sufficiently many states available in one mode, as is the case for coherent-state mode preparation, then ESD and true long-time revivals are also predicted \cite{Yonac-Eberly08}.

\section*{What are future prospects?}

\textit{Quantum memory banks.}  Clearly, ESD can
be largely ignored, to a first approximation, when desirable quantum
operations can be manipulated at sufficiently high speed. The key
goal of memory is opposite of speed, i.e., to preserve quantum state
features semi-indefinitely. Quantum memory networks will be
sensitive to the consequences if ESD occurs. ESD will probably have
to be taken into account if practical versions of quantum memories
are built to operate in mixed-state configurations.

\textit{Disentanglement control.} Over a given noisy channel, it appears that some entangled states may be more robust against the influence of noise than others. In order to control decoherence optimally, it will be useful to learn how to identify the robust states separately from the fragile ones. Control issues also include the use of external fields to manipulate qubit states \cite{Nha-Carmichael} as in gate operations, to create transient decoherence-free subspaces, as mentioned already. A qualitatively different route to combat decoherence specifically of the ESD type is illustrated in Fig. 2. Two evolution tracks are highlighted to show that for qubits prepared with the same value of initial entanglement their concurrences may evolve very differently. In the illustration the right track is subject to ESD, while the left one is not. Note that decoherence per se is not avoided, because the non-ESD track shows steady dissipation, but it always remains finite. Other examples of this are known and in some cases a purely local operation (i.e., a manipulation of only one of the two entangled qubits) can be undertaken to change the state matrix $\rho$ without changing its degree of entanglement, but in a way that switches the evolution trajectory from ESD to non-ESD \cite{Yu-Eberly07QIC}, effectively putting it on a half-life track as in the figure.  Similar studies \cite{Rau-etal} have examined the effect of local operations at intermediate stages of evolution. Use of this method requires detailed knowledge of the state matrix $\rho$, which may not be practical, particularly at times late in the evolution.

\textit{Entanglement invariants.} Entanglement flow in small reservoirs has led to the recent discovery of entanglement ``invariants"  \cite{Yonac-etal07} by inspection of the curves in the bottom row of Fig. 4, which repeats the top row except that a slightly different cat state is employed for the atoms initially:
\beq \label{e.PsiAtoms}
\Psi_\alpha = [(+-)\cos\alpha  \Leftrightarrow (-+)\sin\alpha ].
\eeq
In the bottom left panel of Fig. 4 it appears that each $AB$ atomic concurrence curve is compensated at all times by the corresponding $ab$ photonic concurrence curve in the right panel, one going up as the other falls.

In fact, exact compensation can be confirmed analytically \cite{Yonac-etal07}, but one notes that the same behavior does not appear in the top-row curves in Fig. 4, where there is no perfect compensation of $ab$ for $AB$.  For example, it is easy to see that the  $AB$ and $ab$ red curves can be zero at the same time. A natural question is where does the missing information go in that case? Since the two-site JC model is unitary, preservation of all 4-qubit information is guaranteed, so it should be located ``somewhere". Careful examination shows that concurrence is not conserved, but rather $Q(t)$ is conserved, spread among all 6 different types in that case \cite{Yonac-etal07}.

This identification of 4-particle memory flow channels is unusual and clearly deserves future examination (see \cite{Sainz-Bjork, Cavalcanti-etal}). One can say about these invariants that they emerge only from a kind of analytic continuation of the bipartite concurrence function $C(t)$ to un-physically negative values, which is permitted via $Q(t)$. The entanglement flow issue \cite{Cubitt-etal08} is also related to, and appears to expand considerably, the theory associated with entanglement swapping, which is under active exploration, and has been realized with particle pairs from independent sources \cite{Genevaswap}.

\textit{Non-Markovian noises.} Dissipative entanglement evolution is critically dependent on the types of the noises acting on the system. Markov environments are those for which a noise signal has no self-correlation over any time interval, and under Markov conditions noise typically results in a quantum irreversible process. Non-Markovian noise arising from a structured environment or from strong coupling appears more fundamental (see \cite{nonMarkov1, nonMarkov2}). Recent studies have suggested that correlated noises may cause new difficulties in employing quantum error correction codes \cite{Correlatednoise2} and dynamic decoupling technique \cite{Shiokawa-Hu07}. Although some progress has been made, it is a challenge for the future to extend the current research on ESD into physically relevant non-Markovian situations. High-Q cavity QED and quantum dot systems are two possible experimental venues.

\textit{Qutrits and beyond.} Many-qubit entanglement and entanglement of quantum systems that are not qubits, i.e., those having more than two states, is largely an open question, and one that is embarrassing in a sense, since the question has been open since quantum mechanics was invented in the 1920's. There is still no known finite algorithm for answering the simple-seeming question whether a given mixed state is entangled or not, if it refers to more than two systems, and it is answerable for two mixed-state systems only in the case of pairs of qubits (as we've been discussing) and the case of one qubit and one qutrit (a three-state system such as spin-1).  Investigation into ESD of qubit-qutrit systems has begun \cite{Ann-Jaeger, qutrit2}, but generalizations to many-qubit systems are daunting tasks due to both technical and conceptual difficulties \cite{Simon-Kempe02, Dur-Briegel04,report}.

\textit{Topological approach.} N-party entanglement dynamics will presumably become simpler to predict if a computable entanglement measure for a mixed state of more than two qubits can be discovered. However, an alternative approach is to avoid dynamics through topological analysis \cite{Fine-etal05}. One now knows \cite{Yu-Eberly07warm, AlQasimi-James} that ESD is necessary (i.e., must occur) in arbitrary N-party systems of  non-interacting qubits if they are exposed to thermal noise at any finite $T > 0$ temperature. The steady state of any non-interacting N-qubit system has a neighborhood in which every state is separable. In this case, any prearranged subsystem entanglement will inevitably be destroyed in a finite time. This is a universal result showing how entanglement evolves in the absence of external noise control.

Clearly, the holy grail for research on entanglement dynamics is to
find an efficient real-time technique for tracking and controlling
the entanglement evolution of a generic many-qubit system. Another important open question is to determine a generic method for direct experimental registration of entanglement, for which there is no current answer. We expect many surprising results to be awaiting discovery.


\begin{thebibliography}{99}
\bibitem{SOM2} 
Matrix encoding of the entanglement is more complicated than the bracket notation used for the cats, but is important because we want to allow for more possibilities, which can be visualized by thinking that any of the four joint state labels $++,\ +-,\ -+,\ --$ can be correlated with itself or any other, making 16 basic correlation labels, and these are conventionally arranged in a $4 \times 4$ matrix. This is the density matrix $\rho$ referred to in the text. See Supporting Online Material for an example, and a sketch of the theory of time evolution of the density matrix in a noisy environment.
\end{thebibliography}
\end{document}